\documentclass[a4paper,11pt]{article}
\pdfoutput=1 
\usepackage{changepage}
\usepackage{caption}
\usepackage{color}
\usepackage{amsmath,amssymb}
\usepackage{graphicx}
\usepackage{float}

\usepackage[T1]{fontenc} 

\newcommand{\be}{\begin{eqnarray}}
	\newcommand{\ee}{\end{eqnarray}}

\usepackage{jcappub} 

\usepackage[T1]{fontenc} 
\usepackage{float}
\usepackage{braket}
\usepackage{graphicx}
\usepackage{amsmath}
\usepackage{amsfonts}
\usepackage{amssymb}
\usepackage[english=british]{csquotes}
\usepackage[font=small,labelfont=bf]{caption}
\title{\boldmath Small noise expansion of stochastic inflation}

\author[a]{Diego Cruces,} 
\affiliation[a]{CAS Key Laboratory of Theoretical Physics , Institute of Theoretical Physics, Chinese Academy of Sciences, \\ Beijing, 100190, China}
\author[b]{Cristiano Germani,}
\affiliation[b]{Universitat de Barcelona and Institut de Ci\`encies del Cosmos (ICCUB), \\
	Mart\' i Franqu\`es 1, 08028 Barcelona, Spain }
\author[c]{Amin Nassiri-Rad}
\affiliation[c]{Department of physics, K.N Toosi University of Technology, \\ P.O. Box 15875-4416, Tehran, Iran}
\author[d,e]{and Masahide Yamaguchi}
\affiliation[d]{Cosmology, Gravity and Astroparticle Physics Group, Center for Theoretical Physics of the Universe, Institute for Basic Sclence (IBS), \\ Daejeon, 34126, Korea}

\affiliation[e]{Department of Physics, Institute of Science Tokyo \\ Tokyo, 152-8551, Japan}

\emailAdd{dcruces@itp.ac.cn}
\emailAdd{amin.nassiriraad@kntu.ac.ir}
\emailAdd{germani@icc.ub.edu}
\emailAdd{gucci@ibs.re.kr}

\abstract{By introducing the small noise expansion techniques, we show that the fully non-linear (non-Markovian) stochastic inflationary system, may be re-cast in terms of an infinite set of Wiener processes (stochastic equations with white noises). As a byproduct, we show that the Starobinsky test field approximation might only provide information about the linear regime of cosmological perturbations and, scalar-field non-Gaussianities might only appear at leading order in slow-roll parameters.}

\begin{document}

	\maketitle
	\flushbottom
	
	\section{Introduction}
\label{sec:intro}

To date, one of the main unresolved mysteries that our universe hides is the nature of Dark Matter. Because Dark Matter may only interact gravitationally, long ago \cite{hawking, Chapline:1975ojl}, it was hypothesized that it might be fully made by black holes. However, the recent prolific astronomical search of black holes led to severe limits to this idea \cite{Muller:2024pwn, Carr:2020gox, Carr:2021bzv}. Nevertheless, hypothetical sub-solar black holes might still account for the full of Dark Matter. Because of the Chandrasekhar limit \cite{Chandrasekhar:1931ih}, those black holes cannot be related to astrophysical processes, thus, a more exotic physics must be invoked.  

On the other hand, to explain the cosmic microwave background anisotropies, to date, the most compelling mechanism is inflation \cite{Guth:1980zm,Mukhanov:1981xt,Kodama:1984ziu,Mukhanov:1988jd}. In its simplest version, inflation is a rapid expansion of the primordial universe (well before structure formation and nucleosynthesis) sourced by a scalar field, the inflaton. Quantum fluctuations, classicalizing at super-(cosmological)horizon scales \cite{Polarski:1995jg,Lesgourgues:1996jc}, are those that later produce the observed fluctuation in the temperature of the last scattering photons (those free to propagate in the universe).

A minimal scenario would then be to connect inflation to black hole`s formation. Because inflation generates perturbation quantum mechanically, there is a statistical chance to generate large fluctuation collapsing into black holes of sizes that are not restricted by the Chandrasekhar limit. Although those are rare events (inflationary perturbations are mostly, statistically, small) it might have been enough to have an extremely small density of black holes at formation if they were generated during the radiation epoch of our universe. The reason is that black holes, only interacting gravitationally, are diluted way slower than radiation. Thus, even an extremely small fraction of black holes generated during the radiation epoch can dominate the universe matter content today. These black holes are commonly called Primordial Black Holes (PBH), (see \cite{Escriva:2022duf} for a review). 

This scenario arguably requires the knowledge of the statistics of cosmological perturbations in their non-perturbative regime (see e.g. \cite{Germani:2019zez}).

Before classicalizing (i.e. when the wavelength of the perturbations is small compared to the Hubble scale), the cosmological perturbations are purely quantum and so microscopic. In this regime, the Universe expansion plays a little role. It is only after crossing the horizon (wavelength are stretched in time) that a large number of particle is generated due to the Universe expansion and a sort of cosmic condensate of perturbations is formed with macroscopic energy-density. This is the essence of the stochastic approach to inflation. Macroscopic super-horizon perturbations can be generated as a cumulative effect of many perturbative quantum states crossing continuously (in time) the horizon. 

Those macroscopic fluctuations are causally disconnected with a universe observer until re-entering the horizon, where, some of those, might collapse into black holes if ``large'' enough. 

By definition of super-horizon scales, the classicalized fluctuations have ``small'' gradients with respect to the Hubble scale. In this sense, while sub-horizon quantum states (or ultra-violet states UV) can be treated perturbatively, the super-horizon condensate (the infra-red modes IR) can be large in amplitude but small in gradients, thus be expanded on small spatial derivatives (gradient expansion), see e.g. \cite{Parry:1993mw,Nambu:1994hu,Soda:1995fz, Leach:2001zf,  Tanaka:2006zp, Tanaka:2007gh}. 

Stochastic inflation \cite{Starobinsky:1986fx, Nambu:1987ef, Nambu:1988je, Kandrup:1988sc, Nambu:1989uf, Mollerach:1990zf, Linde:1993xx, Casini:1998wr, Finelli:2008zg, Finelli:2010sh, Assadullahi:2016gkk, Pattison:2017mbe, Firouzjahi:2018vet, Cruces:2018cvq, Prokopec:2019srf, Ballesteros:2020sre, Pattison:2021oen,Clesse:2015wea, Tada:2021zzj,  Cruces:2022imf, Cruces:2021iwq, Tomberg:2023kli, Vennin:2024yzl, Mizuguchi:2024kbl}  connects the two realms by separating the cosmological perturbations into large IR modes sourced by small UV states acting like stochastic kicks (noises).

While the stochastic kicks, at the perturbative level, are approximately Gaussian, they depend on the IR background, making the system not Markovian. This has limited the applicability of the stochastic inflation idea to the case in which the noises are calculated over the deterministic background or, at most, to typical trajectories\footnote{By typical trajectory, we mean a trajectory whose statistical deviation from the background trajectory is small. For example, a typical trajectory when the background is slow-rolling also follows the attractor SR solution.} at leading order in slow-roll parameters, as done in \cite{Vennin:2015hra} where also leading non-gaussianities were obtained. In this respect however, stochastic inflation might only bring perturbative information about cosmological inhomogeneities \cite{Cruces:2021iwq}.

In this paper we go beyond the Markovian approximation. By requiring the physical condition that the stochastic kicks are perturbatively small in any IR background, we show that the stochastic inflationary system might be solved as an infinite set of Markovian (Wiener) processes.       

\section{The small noise expansion}\label{sec:small_noise}
As an illustrative example of what we are going to use in the more complicated stochastic inflation system, we will study the following one-dimensional example (for more details see e.g. \cite{gardiner1985hsm})

\begin{equation}
	d y= a(y)dt + b(y) d\bar W(t)\,,
	\label{small_noise_ex_bar}
\end{equation}
where $\bar W(t)$ represents a Wiener process i.e. $\frac{d\bar W(t)}{dt}=\bar\xi(t)$ with variance $\langle\bar\xi(t)\bar\xi(t')\rangle=\lambda^2\delta(t-t')$. We suppose now that $\lambda\ll 1$ and we recast \eqref{small_noise_ex_bar} into
\begin{equation}
	d y= a(y)dt + \lambda b(y) d W(t)\,,
	\label{small_noise_ex}
\end{equation}
where $\frac{d W(t)}{dt}=\xi(t)$ with variance $\langle\xi(t)\xi(t')\rangle=\delta(t-t')$.
Defining
\begin{equation}
	y(t)\equiv y_0(t)+\sum_{k=1}^\infty\lambda^k y_k(t)\ ,\label{series}
\end{equation}
it can be shown \cite{gardiner1985hsm} that if $\lambda b(y_0)\ll 1$, under quite generic analytic conditions of $b(y)$, the functions $y_k(t)$ solutions of the expanded system \eqref{small_noise_ex} in power of $\lambda$, satisfy the asymptotic condition
\be
y(t)-\sum_{k=0}^n\lambda^k y_k(t)\sim \lambda^{n+1}\ , 
\ee 
for any $y(t)$, in the limit of large $n$.

In other words, although the solution of \eqref{small_noise_ex} is obtained as a series in $\lambda$, it is valid for all amplitudes of $y(t)$, $a(y)$ and $b(y)$. This resonates quite well with our stochastic system: while the kicks can statistically be treated perturbatively (the variance is small and so the chances of a large kick is negligible) the IR modes might be large due to cumulative effects in time.

With this in mind, we can now solve \eqref{small_noise_ex} order by order. First of all we shall expand the function $a(y)$:
\begin{align} \nonumber
	a(y)= & a\left(y_0(t)+\sum_{m=1}^{\infty}\lambda^m y_m \right)=\sum_{p=0}^{\infty} \frac{1}{p!}\frac{d^p a(y_0)}{d y_0^p}\left(\sum_{m=1}^{\infty}\lambda^m y_m \right)^p \\
	& a(y_0) + \lambda \left(y_1\frac{d a(y_0)}{d y_0}\right) + \lambda^2 \left(y_2\frac{d a(y_0)}{d y_0}+\frac{1}{2}y_1^2\frac{d^2 a(y_0)}{d y_0^2}\right)+\ldots \,.
	\label{a_exp}
\end{align}

Although it is not easy to write explicitly the full set of terms in general, it is easy to see that, for $n \geq 1$, the one proportional to $\lambda^n$ will only linearly depend on $y_n$. Inserting the expansion \eqref{a_exp} for $a(y)$ and an equivalent one for $b(y)$ into \eqref{small_noise_ex}, we can solve the equations by equating coefficients of like powers of $\lambda$. We then obtain an infinite set of stochastic differential equations:

\begin{align} \nonumber
	d y_0 & = a(y_0) dt \\ \nonumber
	d y_1 & = \frac{d a(y_0)}{d y_0} y_1 dt + b(y_0)dW(t) \\ \nonumber
	d y_2 & = \left(\frac{d a(y_0)}{d y_0} y_2 + \frac{1}{2}\frac{d^2 a(y_0)}{dy_0^2} y_1^2\right) dt + \frac{d b(y_0)}{d y_0} y_1 dW(t) \\ 
	& \vdots \hspace{9cm}\,.
	\label{small_noise}
\end{align}

From \eqref{small_noise} we can easily see that the probability distribution for $y_1$ obtained when solving the equation at leading order in the small noise expansion, is exactly Gaussian\footnote{In fact,  as also noted in \cite{Ballesteros:2020sre}, one can show that the first non-Gaussian contribution appears in the equation for $y_2$ such that 
	$$\langle \left(y(t) - \langle y(t)\rangle\right)^3 \rangle =  3\left(\langle y_1^2 y_2\rangle-\langle y_1^2\rangle\langle y_2\rangle\right)\lambda^4 + \mathcal{O}\left(\lambda^5\right)$$ \label{foot:non_gaussian}}. This is because $W(t)$ is a Weiner process and the equation for $y_1$ is linear in the stochastic variable.

It is also important to note that if $b(y)$ depends generically on $y$, the approximation $b(y)\simeq b(y_0)$, where $y_0$ is the deterministic value of $y$ (or zeroth order in small noise expansion), enters in the small noise expansion at leading order $\lambda$. This is what is typically done in stochastic inflation in order to approximate the full system into a Wiener process (i.e. a process with white noises).

\section{The gradient expansion and the on-set of stochastic inflation}

Suppose to have a perturbation of instantaneous wavelength $L$. Although the Universe is generically inhomogeneous, one might consider a small almost homogeneous and isotropic patch of local Hubble scale $H^{IR}$. We say that the given perturbation is super-horizon whenever $L H^{IR}\gg 1$. Because we will deal with derivatives (which are the inverse of the wavelengths) we shall introduce a parameter $\sigma\equiv (L H^{IR})^{-1}\ll 1$\footnote{The transition between IR and UV modes is obviously not sharp. Moreover, the cosmological horizon is $H^{IR}$ only at leading order in slow roll parameters. The constant $\sigma$ is then conservatively introduced to parameterize our ignorance. In other words, modes of wavelengths well above $L_c\equiv(\sigma H^{IR})^{-1}$ are surely casually disconnected from modes much smaller than $L_c$.}.

Having in mind a background Friedman-Robertson-Walker (FRW) metric, a generic perturbation can be decomposed following the ADM formalism \cite{Arnowitt:1959ah}
\be
ds^2=-\alpha^2 dt^2+a(t)^2 e^{2\zeta}\tilde \gamma_{ij}(dx^i+\beta^i dt)(dx^j+\beta^j dt)\ ,
\label{ADM_metric}
\ee
where $a(t)$ is the background scale factor written in cosmic time $t$. The background FRW universe in its explicit isotropic form, is obtained for $\alpha=0$, $\zeta=0$, $\tilde\gamma_{ij}=\delta_{ij}$ and $\beta^i=0$.

Expanding the metric components in gradients, by the use of Einstein equations, one generically finds that, for any local patch and at leading order in $\sigma$ \cite{Cruces:2021iwq}\footnote{This is indeed generic, however, there are some specific gauges that do not respect those orders \cite{Cruces:2022dom}. The most common one is the Newtonian gauge. Nevertheless, in the following, we shall use the well behave uniform $N$ gauge.}
\be
\alpha={\cal O}(\sigma^0)\ ,\tilde\gamma_{ij}-\delta_{ij}={\cal O}(\sigma)\ ,\zeta={\cal O}(\sigma^0)\ ,\beta^i={\cal O}(\sigma^{-1})\ .
\ee
This is compatible with the fact that locally, at leading order in $\sigma$, the universe is FRW. It might be though surprising to see that the shift $\beta^i$ is ${\cal O}(\sigma^{-1})$. However, a local FRW universe might be also written in a non-isotropic form where $\beta^i\propto x^i$ \cite{Cruces:2021iwq}.

Finally, the special metric hides a zeroth order term. In fact, one may rewrite
\be
\tilde\gamma_{ij}-\delta_{ij}=-2 \left(\partial_i\partial_j-\frac{1}{3}\delta_{ij}\nabla^2\right)C+{\cal O}(\sigma^2)\ ,
\ee 
with $\nabla^2 C={\cal O}(\sigma^0)$. The component $C$ is related to the momentum constraint of the Einstein equations and cannot be neglected at next to leading order in the slow-roll parameter $\epsilon_1\equiv-\frac{\dot H}{H^2}$ \cite{Cruces:2021iwq} as routinely done in the literature \cite{Langlois:2005ii, Rigopoulos:2004gr}. We shall discuss this later on in relation to the small noise expansion. 

For convenience, we will formulate stochastic inflation in the uniform-$N$ gauge where the number of e-folds ($N$) of inflation are calculated on the background\footnote{By using the metric \eqref{ADM_metric}, $N$ is defined as
	$$N\equiv\int \left(\bar{H} + \dot{\zeta} - \frac{D_i\beta^i}{3}\right)dt\,,$$ 
	where $D_i$ is the covariant derivative with respect to the metric $\gamma_{ij}=e^{2\zeta}\tilde{\gamma}_{ij}$ and $\bar{H}=\frac{\dot{a}}{a}$ is the Hubble parameter in the FRW background universe, where the number of e-folds is straightforwardly given by
	$$\bar{N} = \int \bar{H} dt\,,$$ the uniform $N$ gauge is such that $\bar{N}=N$.}, where $\zeta=0$ and $\beta^i=0$. Due to the usage of this gauge, we will use $N$ as the time variable from now on. 

Having set the order of the IR metric functions, in stochastic inflation, each term is split into UV and IR modes (see \cite{Launay:2024qsm} for a more physical version of this) such that, for a generic function X, we have the following decomposition

\begin{align} \nonumber
	X^{IR}(N,\textbf{x})\equiv \int \frac{d\textbf{k}}{(2\pi)^3}\Theta\left(\sigma a (N)H^{IR}(N)-k\right)\hat{\mathcal{X}}^{IR}_{\textbf{k}}(N,\textbf{x})\,, \\
	X^{UV}(N,\textbf{x})\equiv \int \frac{d\textbf{k}}{(2\pi)^3}\Theta \left(k-\sigma a(N)H^{IR}(N)\right)\hat{\mathcal{X}}^{UV}_{\textbf{k}}(N,\textbf{x})\,.
	\label{IR_UV_def}
\end{align}

Similarly as in linear perturbation theory $\hat{\mathcal{X}}^{UV}_{\textbf{k}}(t,\textbf{x})$ is defined as the following hermitian operator:

\begin{equation}
	\hat{\mathcal{X}}^{UV}_{\textbf{k}}(N,\textbf{x})=X^{UV}_{\textbf{k}}(N)\hat{a}_{\textbf{k}}e^{i\mathbf{k}\cdot \mathbf{x}} + \left(X^{UV}_{\textbf{k}}\right)^{\star}(N)\hat{a}_{\textbf{k}}^{\dagger}e^{-i \mathbf{k}\cdot \mathbf{x}}\,,
	\label{X_decomposition}
\end{equation}
where $X^{UV}_{\textbf{k}}(N)$ is the solution of the evolution equation for the perturbation $X^{UV}$ over the local background. The operators $a_{\textbf{k}}$ and $a^{\dagger}_{\textbf{k}}$ are the usual creation and annihilation operators.

Note that, in the spirit of gradient expansion, the splitting is done in the local cosmological coarse-grained scale $(\sigma H^{IR})^{-1}$, which generically differs form the one of the background, for example in uniform-N gauge we have $H^{IR}=\frac{\bar{H}}{\alpha^{IR}}$.

After expanding the ADM equations at leading order in gradient expansion for the IR variables and at leading order in perturbation theory for the UV variables. Inserting the decomposition \eqref{X_decomposition} for the UV variables, the final equations in the uniform-N gauge read (see \cite{Cruces:2022imf} for a detailed derivation):

\begin{align} \nonumber
	\pi^{IR}&=\frac{\partial\phi^{IR}}{\partial N}+\xi_{\phi}\,,\\ \nonumber
	\frac{\partial\pi^{IR}}{\partial N}&+\left(3-\frac{\left(\pi^{IR}\right)^2}{2M_{PL}^2}\right)\pi^{IR}+\left(3M_{PL}^2-\frac{\left(\pi^{IR}\right)^2}{2}\right)\frac{V_{\phi}\left(\phi^{IR}\right)}{V\left(\phi^{IR}\right)}=-\xi_{\pi}\,,\\
	\frac{\partial }{\partial N}&\left(\frac{1}{3}\nabla^2 C^{IR}\right)-\left(\frac{\bar{H}}{H^{IR}}-1\right)+\frac{1}{2M_{PL}^2}\frac{\partial\phi^{IR}}{\partial N}\left(\phi^{IR}-\bar{\phi}\right)=-\xi_C\,. 
	\label{full_stochastic_system}
\end{align}
where $$H^{IR}=\sqrt{\frac{V\left(\phi^{IR}\right)}{3M_{PL}^2-\frac{\left(\pi^{IR}\right)^2}{2}}}\,,$$
and the noises are specified as follows
\begin{align} \nonumber
	\langle\xi_\phi(N_1)\xi_\phi(N_2)\rangle&=\left(1+\frac{\partial \ln H^{IR}}{\partial N}\right)  P^2_\phi\delta (N_1-N_2)\,, \\ \nonumber
	\langle\xi_\pi(N_1)\xi_\pi(N_2)\rangle&=\left(1+\frac{\partial \ln H^{IR}}{\partial N}\right) P^2_\pi\delta (N_1-N_2)\,, \\
	\langle\xi_C(N_1)\xi_C(N_2)\rangle&=\left(1+\frac{\partial \ln H^{IR}}{\partial N}\right) P^2_C\delta (N_1-N_2)\,. 
	\label{noises}
\end{align}
The dirac delta appearing in \eqref{noises} is a direct consequence of the use of the Heaviside theta function as the window function responsible for the splitting between IR and UV modes. We have already mentioned that this choice is rather unphysical \cite{Winitzki:1999ve} as it assumes that the UV modes instantaneously lose their quantum nature to convert into IR modes. In principle this can be refined \cite{Mahbub:2022osb}, however, it is beyond the scope of this paper. 

Finally, in \eqref{noises}, we have defined the square root of the stochastic power spectrum as follows
\begin{align} \nonumber
	P_{\phi}&\equiv \sqrt{\left(\frac{k^3}{2\pi^2}\left|\phi^{UV}_k\right|^2\right)_{k=\sigma a H^{IR}}}\,, \\ \nonumber
	P_{\pi}&\equiv \sqrt{\left(\frac{k^3}{2\pi^2}\left|\pi^{UV}_k\right|^2\right)_{k=\sigma a H^{IR}}}\,, \\
	P_{C}&\equiv \sqrt{\left(\frac{k^3}{2\pi^2}\left|k^2C^{UV}_k\right|^2\right)_{k=\sigma a H^{IR}}}\,,
	\label{PS_stochastic}
\end{align}
where the UV perturbations, in the uniform-N gauge, are 
\begin{equation}
	\phi^{UV}_k=-\frac{\partial \phi^{IR}}{\partial N}\mathcal{R}^{sto}_k-\frac{1}{3}k^2C^{UV}_k\,, \qquad \pi^{UV}_k=\frac{\partial \phi^{UV}_k}{\partial N}\,.
\end{equation}
The function $\mathcal{R}^{sto}_k$ is the gauge invariant comoving curvature perturbation solving the Mukhanov-Sasaki (MS) equation over a stochastic background\footnote{We define here $\epsilon_i^{IR}$ in a similar way as its deterministic counterpart i.e.
	$$\epsilon_1^{IR}=-\frac{\partial \ln H^{IR}}{\partial N}\,, \qquad \epsilon_{i+1}^{IR}=\frac{\partial \ln \epsilon_i^{IR}}{\partial N} \quad \text{for} \quad i>1\,. $$}:
\begin{equation}
	\frac{H^{IR}}{a^3\epsilon_1^{IR}}\frac{d}{d N}\left(a^3\epsilon_1^{IR} H^{IR}\frac{d \mathcal{R}^{sto}_k}{d N}\right)+k^2\mathcal{R}^{sto}_k=0\, ,
	\label{MS_equation_com_sto}
\end{equation}
while the equation for $C^{UV}_k$ is 
\begin{equation}
	\frac{\partial}{\partial N} \left(\frac{1}{3}k^2 C_k^{UV}\right) -\frac{k^2}{\left(H^{IR}\right)^2 a^2}\left(\frac{1}{3} k^2 C_k^{UV}\right) = \frac{\epsilon_1^{IR}}{(3-\epsilon_1^{IR})\pi^{IR}}\pi_k^{UV}-\frac{\epsilon_1^{IR}\epsilon_2^{IR}}{2\left(3-\epsilon_1^{IR}\right)\pi^{IR}}\phi_k^{UV}\,.
	\label{C_sto}
\end{equation}

\section{Small noise expansion in stochastic inflation}

The concept of small noise expansion in the context of stochastic inflation has been used before at next to leading order \cite{Kunze:2006tu, Ballesteros:2020sre}. There, it was used to consider the leading backreaction of the stochastic field to the evolving background in the perturbative regime. Here, for the first time, we present a systematic derivation of the stochastic equations at all orders in small noises. As we already pointed out, as long as the variance of the noises is small, the series in noises captures the full non-perturbative behavior of the non-linear stochastic variable. Thus, this analysis is essential to get information over the tail of its probability distribution, which is specially relevant for PBH formation

In order to make a clear connection with the small noise expansion presented in section \ref{sec:small_noise} we will rewrite the system \eqref{full_stochastic_system} in the following way

\begin{align} \nonumber
	\pi^{IR}&=\frac{\partial\phi^{IR}}{\partial N}+\lambda^i_\phi\,  b_{\phi}\, \xi\,,\\ \nonumber
	\frac{\partial\pi^{IR}}{\partial N}&+\left(3-\frac{\left(\pi^{IR}\right)^2}{2M_{PL}^2}\right)\pi^{IR}+\left(3M_{PL}^2-\frac{\left(\pi^{IR}\right)^2}{2}\right)\frac{V_{\phi}\left(\phi^{IR}\right)}{V\left(\phi^{IR}\right)}=-\lambda^i_\pi\,  b_{\pi}\, \xi\,,\\
	\frac{\partial }{\partial N}&\left(\frac{1}{3}\nabla^2 C^{IR}\right)-\left(\frac{\bar{H}}{H^{IR}}-1\right)+\frac{1}{2M_{PL}^2}\frac{\partial\phi^{IR}}{\partial N}\left(\phi^{IR}-\bar{\phi}\right)=-\lambda^i_C\,  b_{C}\, \xi\,. 
	\label{full_stochastic_system2}
\end{align}
where  $\langle\xi(N_1)\xi(N_2)\rangle=\delta(N_1-N_2)$ and 
\begin{align} \nonumber
	\lambda_{\phi}&=\sqrt{\left(\frac{k^3}{2\pi^2}\left|\bar{\phi}^{UV}_k\right|^2\right)\Big|_{k=\sigma a \bar H}}\,, \\ \nonumber
	\lambda_{\pi}&=\sqrt{\left(\frac{k^3}{2\pi^2}\left|\bar{\pi}^{UV}_k\right|^2\right)\Big|_{k=\sigma a \bar H}}\,, \\
	\lambda_{C}&=\sqrt{\left(\frac{k^3}{2\pi^2}\left|k^2\bar{C}^{UV}_k\right|^2\right)\Big|_{k=\sigma a \bar H}}\, ,
\end{align}
are the square root of the UV power spectrum on the deterministic background\footnote{This power spectrum is simply computed by solving \eqref{MS_equation_com_sto} and \eqref{C_sto} but over a deterministic background (changing the $IR$ variables by those of the fiducial background).}.

Finally, we defined
\begin{align} \nonumber
	b_{\phi}&\equiv \left(1+\frac{\partial \ln H^{IR}}{\partial N}\right)^{1/2} \frac{ P_{\phi}}{\lambda^i_\phi}\,, \\ \nonumber
	b_{\pi}&\equiv \pm\left(1+\frac{\partial \ln H^{IR}}{\partial N}\right)^{1/2} \frac{ P_{\pi}}{\lambda^i_\pi}\,, \\ 
	b_{C}&\equiv \pm\left(1+\frac{\partial \ln H^{IR}}{\partial N}\right)^{1/2} \frac{ P_{C}}{\lambda^i_C}\ ,
	\label{definition_b}
\end{align}
and $\lambda^i\equiv\lambda(t=t_i)$, where $t_i$ is the initial time of inflation. This, corresponds to when the first mode ($k=\sigma a(t_i) H(t_i)$) of interest crosses the coarse-grained scale. For the rest of the paper we will assume, without loss of generality, that this happens at $N=0$. Note that in \eqref{full_stochastic_system2}, $\lambda^i$ acts as the expansion parameter and it extracts the order of magnitude of the noises such that $b$ is $\mathcal{O}(1)$ whenever evaluated in the background. Finally, the sign in the second and third lines of \eqref{definition_b} depends on the sign of the cross-correlators $\langle\xi^{\lambda}_{\phi}(N_1)\xi^{\lambda}_{\pi}(N_2)\rangle$ and $\langle\xi^{\lambda}_{\phi}(N_1)\xi^{\lambda}_{C}(N_2)\rangle$, respectively. 

Because the $\lambda^i$´s are small, we could use them as a small parameter as long as 
\be 
\lambda^i b\Big|_{\rm background}\equiv \lambda^i \bar{b} = \sqrt{1-\epsilon_1}\lambda \ll 1\ .
\label{condition_expansion}
\ee 
The condition \eqref{condition_expansion} is going to be always satisfied in the cases of our interest. One might be however worried about cases in which the ${\cal O}(\lambda^0)$ solution for \eqref{full_stochastic_system2} becomes smaller than the one at leading order ${\cal O}(\lambda^1)$. This is what sometimes has been dubbed diffusion dominated regime in the context of stochastic inflation \cite{Pattison:2021oen}. As shown in section 6.2.4 of \cite{gardiner1985hsm}, a simple change of variables still allows to use consistently the small noise expansion even in those degenerate limits. In the following, in order to avoid this technicality, without any loss of generality, we will assume that we can always perform the small noise expansion in the original variables.

In the last part of this section we shall present the stochastic equations up to ${\cal O}(\lambda^2)$:

\begin{itemize}
	\item At zeroth order in the small noise expansion $(\lambda^0)$, we obtain the background equations (where $\left(\phi_0,\pi_0\right)=\left(\bar{\phi},\bar{\pi}\right)$):
	\begin{align} \nonumber
		\bar{\pi}= & \frac{\partial\bar{\phi}}{\partial N}\,,\\ 
		\frac{\partial\bar{\pi}}{\partial N} & +\left(3-\frac{\bar{\pi}^2}{2M_{PL}^2}\right)\bar{\pi}+\left(3M_{PL}^2-\frac{\bar{\pi}^2}{2}\right)\frac{V_{\bar{\phi}}\left(\bar{\phi}\right)}{V\left(\bar{\phi}\right)}=0\,,
		\label{small_noise_0_inf}
	\end{align}
	
	\item At leading order in the small noise expansion $(\lambda^1)$, we have:
	\begin{align} \nonumber
		\frac{\partial \phi_1}{\partial N} = & \pi_1 + \bar{b}_{\phi}\xi\,, \\ \nonumber
		\frac{\partial \pi_1}{\partial N}  + & \left(3 - \epsilon_1 + \frac{\epsilon_1\epsilon_2}{3-\epsilon_1}\right)  \pi_1 + \left(-\frac{3}{2}\epsilon_2+\frac{1}{2}\epsilon_1\epsilon_2-\frac{1}{4}\epsilon_2^2-\frac{1}{2}\epsilon_2\epsilon_3-\frac{\epsilon_1\epsilon_2^2}{2(3-\epsilon_1)}\right)\phi_1 = -\bar{b}_{\pi}\xi\,,
		\label{sto_any_epsilon0_noise1}
		\\ 
		\frac{\partial}{\partial N}& \left(\frac{1}{3}\nabla^2 C_1\right)  + \frac{\epsilon_1}{(3-\epsilon_1)\bar{\pi}}\pi_1-\frac{\epsilon_1\epsilon_2}{2(3-\epsilon_1)\bar{\pi}}\phi_1=-\bar{b}_{C}\xi\,,
	\end{align}
	\item Second order in small noise approximation ($\lambda^2$). In this case the system of equations that we obtain is more involved: 
	\begin{align} 
		\frac{\partial \phi_2}{\partial N} = & \pi_2 + \left(\phi_1\frac{\partial \bar{b}_{\phi}}{\partial \bar{\phi}} + \pi_1\frac{\partial \bar{b}_{\phi}}{\partial \bar{\pi}}\right)\xi\,, \\ \nonumber
		\frac{\partial \pi_2}{\partial N}  + & \left(3 - \epsilon_1 + \frac{\epsilon_1\epsilon_2}{3-\epsilon_1}\right)  \pi_2 + \left(-\frac{3}{2}\epsilon_2+\frac{1}{2}\epsilon_1\epsilon_2-\frac{1}{4}\epsilon_2^2-\frac{1}{2}\epsilon_2\epsilon_3-\frac{\epsilon_1\epsilon_2^2}{2(3-\epsilon_1)}\right)\phi_2 \\ \nonumber
		-\frac{\epsilon_1}{\bar{\pi}} & \left(4-\frac{\epsilon_2}{(3-\epsilon_1)}\right)\pi_1^2 + \frac{\epsilon_1}{\bar{\pi}}\left(\epsilon_2 + \frac{\epsilon_2^2}{2}\frac{(3+\epsilon_1)}{(3-\epsilon_1)^2} + \frac{\epsilon_1\epsilon_2}{(3-\epsilon_1)}\right)\pi_1\phi_1 \\ \nonumber
		-\frac{\epsilon_2}{4\bar{\pi}}&\left(\epsilon_1\epsilon_2^2\frac{9+\epsilon_1}{(3-\epsilon_1)^2}+2\epsilon_2\epsilon_3\frac{3+2\epsilon_1}{3-\epsilon_1}+2\epsilon_3(3-\epsilon_1)+2\epsilon_3^2 + 2\epsilon_3\epsilon_4\right) \phi_1^2 \\ 
		=  - &\left(\phi_1\frac{\partial \bar{b}_{\pi}}{\partial \bar{\phi}} +\pi_1\frac{\partial \bar{b}_{\pi}}{\partial \bar{\pi}}\right)\xi\,,
		\\ \nonumber
		\frac{\partial}{\partial N} & \left(\frac{1}{3}\nabla^2 C_2\right)  + \frac{\epsilon_1}{(3-\epsilon_1)\bar{\pi}}\pi_2-\frac{\epsilon_1\epsilon_2}{2(3-\epsilon_1)\bar{\pi}}\phi_2 \\ \nonumber
		+ &\frac{3}{4 M_{PL}^2(3-\epsilon_1)^2}\pi_1^2 + \frac{18 - \epsilon_1(6-\epsilon_2)}{4M_{PL}^2(3-\epsilon_1)^2}\phi_1\pi_1 \\ \nonumber
		- &\frac{4\epsilon_1^3 - 2\epsilon_1^2 (12 +\epsilon_2)+3\epsilon_2(6+\epsilon_2+2\epsilon_3)+2\epsilon_1(18+\epsilon_2^2-\epsilon_2\epsilon_3)}{16M_{PL}^2(3-\epsilon_1)^2}\phi_1^2 \\ 
		= & -\left(\phi_1\frac{\partial \bar{b}_{C}}{\partial \bar{\phi}} +\pi_1\frac{\partial \bar{b}_{C}}{\partial \bar{\pi}}\right)\xi
		\,,
	\end{align}
\end{itemize}
with those equations we will now be able to state some generic features of the stochastic inflation system.

The noises $\bar{b}_{i}$ used in the small noise expansion of stochastic inflation are simply the noises \eqref{definition_b} but evaluated over the deterministic background, i.e.
\begin{align} \nonumber
	\bar b_{\phi}&=\sqrt{1-\epsilon_1}\frac{ \lambda_{\phi}}{\lambda^i_{\phi}}\,, \\ \nonumber
	\bar{b}_{\pi}&=\pm \sqrt{1-\epsilon_1}\frac{ \lambda_{\pi}}{\lambda^i_{\pi}}\,, \\
	\bar{b}_{C}&=\pm \sqrt{1-\epsilon_1}\frac{ \lambda_{C}}{\lambda^i_{C}}\,, 
	\label{noises_small}
\end{align}
where $\epsilon_1=-\frac{\dot{\bar{H}}}{\bar{H}^2}$ and $\epsilon_i=\frac{\dot\epsilon_{i-1}}{\bar{H}\epsilon_{i-1}}$, for $i>1$.
\section{Tree-level power spectrum}
\label{sec:tree_level}
In this section, we show that the stochastic two-point correlation of the field fluctuation at leading order in $\lambda$ exactly matches the one calculated from linear perturbation theory.

The system \eqref{sto_any_epsilon0_noise1} can be written in the following suggestive form :
\begin{align} \nonumber
	\frac{\partial Q_1}{\partial N}= & \left(\pi_Q\right)_1 + \bar{b}_{Q} \xi\,, \\ 
	\frac{\partial \left(\pi_Q\right)_1}{\partial N}  + & \left(3 - \epsilon_1\right) \left(\pi_Q\right)_1 + \left(-\frac{3}{2}\epsilon_2+\frac{1}{2}\epsilon_1\epsilon_2-\frac{1}{4}\epsilon_2^2-\frac{1}{2}\epsilon_2\epsilon_3\right) Q_1 = -\bar{b}_{\pi_Q}\xi\,,
	\label{sto_any_epsilon0_noise2}
\end{align}
where we have defined\footnote{Note that this change of variables does not affect the equation at leading order in small noise \eqref{small_noise_0_inf}, as $\bar{C}=0$.} $Q_1 \equiv \phi_1 - \bar{\pi} \left(\frac{1}{3}\nabla^2 C_1\right)$ and
\begin{align} \nonumber
	\bar b_{Q}&=\sqrt{1-\epsilon_1}\frac{ \lambda_{Q}}{\lambda^i_{Q}}\,, \\ \nonumber
	\bar{b}_{\pi_Q}&=\pm\sqrt{1-\epsilon_1}\frac{ \lambda_{\pi_Q}}{\lambda^i_{\pi_Q}}\,, \\
	\lambda_{Q}&=\sqrt{\left(\frac{k^3}{2\pi^2}\left|Q^{UV}_k\right|^2\right)\Big|_{k=\sigma a \bar H}}\,, \\ \nonumber
	\lambda_{\pi_Q}&=\sqrt{\left(\frac{k^3}{2\pi^2}\left|\left(\pi_Q^{UV}\right)_k\right|^2\right)\Big|_{k=\sigma a \bar H}}\,.
\end{align}

The variable $Q^{UV}$ is nothing more than the gauge invariant MS variable, i.e.
\begin{equation}
	\phi^{UV}_k=Q^{UV}_k=-\frac{\partial \bar{\phi}}{\partial N}\mathcal{R}_k\,, \qquad \left(\pi_Q^{UV}\right)_k=\frac{\partial Q^{UV}_k}{\partial N}\,,
\end{equation}
so it solves equations \eqref{MS_equation_com_sto} and \eqref{C_sto} over a deterministic background. These equations might be written in terms of $Q_k$ and $\left(\pi_Q\right)_k$ and take a very similar form as \eqref{sto_any_epsilon0_noise2}:
\begin{align} \nonumber
	\frac{\partial Q_k^{UV}}{\partial N}= & \left(\pi_Q^{UV}\right)_k\,, \\ 
	\frac{\partial \left(\pi_Q^{UV}\right)_k}{\partial N}  + & \left(3 - \epsilon_1\right) \left(\pi_Q^{UV}\right)_k + \left[\left(\frac{k}{a\bar{H}}\right)^2+\left(-\frac{3}{2}\epsilon_2+\frac{1}{2}\epsilon_1\epsilon_2-\frac{1}{4}\epsilon_2^2-\frac{1}{2}\epsilon_2\epsilon_3\right)\right] Q_k^{UV} =0\,.
	\label{MS_equation_Q_det}
\end{align}

The stochastic system \eqref{sto_any_epsilon0_noise2} is a 2-dimensional Ornstein–Uhlenbeck process with time-dependent coefficients for which an exact solution exists:

For convenience, we will use the non-linear version of the gauge invariant field perturbation \cite{Cruces:2021iwq} and its time derivative
\begin{equation}
	Q^{IR}=\phi^{IR}-\bar{\phi} - \frac{\partial \phi^{IR}}{\partial N}\frac{1}{3}\nabla^2 C^{IR}\,,\qquad \pi_Q^{IR}=\frac{d Q^{IR}}{d N}\,,
	\label{def_QIR}
\end{equation}
which, at leading order in the small noise expansion, reads
\begin{equation}
	Q^{IR} = \lambda_{Q}^i Q_1 + \mathcal{O}(\lambda^2)\,, \qquad  \pi_Q^{IR} = \lambda_{\pi_Q}^i \left(\pi_Q\right)_1 + \mathcal{O}(\lambda^2)\,.
\end{equation}

Defining the following matrices:
\begin{equation}
	\mathbf{Q}^{IR} \equiv 
	\begin{bmatrix} \lambda_Q^i Q_1\\ \lambda_{\pi_Q}^i \left(\pi_Q\right)_1 \end{bmatrix}+\mathcal{O}(\lambda^2)\,,
\end{equation}
\begin{equation}
	\mathbf{A} \equiv 
	\begin{bmatrix} 0 & -1\\ -\frac{3}{2}\epsilon_2+\frac{1}{2}\epsilon_1\epsilon_2-\frac{1}{4}\epsilon_2^2-\frac{1}{2}\epsilon_2\epsilon_3 & 3 - \epsilon_1 \end{bmatrix}\,,
\end{equation}

\begin{equation}
	\mathbf{B} \equiv 
	\begin{bmatrix} \lambda_{Q}^i\bar{b}_Q \\ \lambda_{\pi_Q}^i\bar{b}_{\pi_Q} \end{bmatrix}\,,
	\label{def_B}
\end{equation}
we find that the two-point correlator for $\mathbf{Q}^{IR}$, obtained by solving \eqref{sto_any_epsilon0_noise2} for $\left(Q_1, \left(\pi_Q\right)_1 \right)$\footnote{Note that, at leading order in small noise expansion, the two-point correlator for $\mathbf{Q}^{IR}$ coincides with the variance of the probability distribution. The reason is that $
	\mathbf{Q}^{IR}(0) =  
	\begin{bmatrix} 0\\ 0 \end{bmatrix}\,,$ and hence $\langle\mathbf{Q}^{IR}(N)\rangle =
	\begin{bmatrix} 0\\ 0 \end{bmatrix} + \mathcal{O}(\lambda^2)\,.$}, is

\begin{equation}
	\langle \mathbf{Q}^{IR}(N) \left(\mathbf{Q}^{IR}\right)^T(N)\rangle = \int_0^N \exp \left[-\int_{N'}^{N}\mathbf{A}(s)ds\right]\cdot\mathbf{B}(N')\cdot \mathbf{B}^{T}(N') \cdot\exp \left[-\int_{N'}^{N}\mathbf{A}^{T}(s)ds\right]dN'\ .
	\label{stochastic_solution_small}
\end{equation}

In order to compute \textbf{B}, we must evaluate the solution of \eqref{MS_equation_Q_det} at the coarse-grained scale ($k=\sigma a \bar{H}$). Since the coarse-grained scale is way larger than the horizon ($\sigma \ll 1$), we can neglect the term proportional to $\left(\frac{k}{a\bar{H}}\right)^2$ in \eqref{MS_equation_Q_det} getting
\begin{equation}
	\mathbf{Q}_k (N) = \exp \left[-\int_{0}^{N}\mathbf{A}(s)ds\right]\mathbf{C}_k\,,
	\label{MS_solution_Q}
\end{equation}
where we have defined, similarly to the IR case, the following vectors
\begin{equation}
	\mathbf{Q}_k \equiv 
	\begin{bmatrix} Q^{UV}_k\\ \left(\pi_Q^{UV}\right)_k \end{bmatrix}\,,
\end{equation}
\begin{equation}
	\mathbf{C}_k \equiv 
	\begin{bmatrix} c_1(k)\\ c_2(k) \end{bmatrix}\,.
\end{equation}
The $k$-dependent constants $c_1(k)$ and $c_2(k)$ are given by initial conditions\footnote{The initial conditions usually employed comes from the Bunch-Davies vacuum \cite{Bunch:1978yq}, however, this argument is also valid when $c_1(k)$ and $c_2(k)$ are specified by matching the solution of the Mukhanov-Sasaki equation (in terms of Hankel functions) at some transition time $N_t$ between different inflationary regimes (see for example \cite{Byrnes:2018txb, Cai:2018dkf})}.

The dimensionless power spectrum of $Q_k^{UV}$ and $\left(\pi_Q^{UV}\right)_k$, needed to find the $\lambda^i$s, as well as the cross-correlations, can then be recast in the matrix (for simplicity we drop the $UV$ index)
\begin{align} 
	\frac{k^3}{2\pi^2} \mathbf{Q}_k(N)\cdot \mathbf{Q}_k^{T}(N)=\begin{bmatrix} \mathcal{P}_Q(k,N) \equiv \frac{k^3}{2\pi^2}\left|Q_k(N)\right|^2 & \frac{k^3}{2\pi^2}Q_k(N)\left(\pi_Q\right)^{\star}_k(N)\\  \frac{k^3}{2\pi^2}Q^{\star}_k(N)\left(\pi_Q\right)_k(N) & \mathcal{P}_{\pi_Q}(k,N) \equiv \frac{k^3}{2\pi^2}\left|\left(\pi_Q\right)_k(N)\right|^2 \end{bmatrix} \,.
	\label{power_spectrum_def}
\end{align}
so that\footnote{At superhorizon scales we have that $\frac{k^3}{2\pi^2}Q^{\star}_k(N)\left(\pi_Q\right)_k(N)=\frac{k^3}{2\pi^2}Q_k(N)\left(\pi_Q\right)^{\star}_k(N)$ is approximately a real variable, which is a consequence of the quantum to classical transition (or decoherence) of inflationary quantum fluctuations at superhorizon scales. Note that this is a necessary condition in order to formulate stochastic inflation since otherwise quantum effects cannot be described as classical random variables.}
\begin{equation}
	\mathbf{B}\cdot \mathbf{B}^{T} = \frac{\left(\sigma a \bar{H}\right)^3}{2\pi^2}(1-\epsilon_1) \left(\mathbf{Q}_k\cdot \mathbf{Q}_k^{T}\right)_{k=\sigma a \bar{H}}\,.
	\label{relation_B_Q}
\end{equation}

Inserting \eqref{MS_solution_Q}
and \eqref{relation_B_Q} into \eqref{stochastic_solution_small} we get
\begin{align} \nonumber
	\langle \mathbf{Q}^{IR}(N) \left(\mathbf{Q}^T\right)^{IR}(N)\rangle = & \int_0^N  \left(\frac{\left(\sigma a(N') \bar{H}(N')\right)^3}{2\pi^2}\left(1-\epsilon_1(N')\right)\right)\\
	& \exp \left[-\int_{0}^{N}\mathbf{A}(s)ds\right]\cdot\left(\mathbf{C}_k\cdot \mathbf{C}_k^{T}\right)_{k=\sigma a(N')\bar{H}(N') } \cdot\exp \left[-\int_{0}^{N}\mathbf{A}^{T}(s)ds\right]dN'
	\label{stochastic_solution_simplified}
\end{align}
where
\begin{small}
	\begin{equation}
		\left(\mathbf{C}_k\cdot \mathbf{C}_k^{T}\right)_{k=\sigma a(N')\bar{H}(N')} =  
		\begin{bmatrix} \left|c_1\left(\sigma a(N') \bar{H}(N')\right)\right|^2 & c_1\left(\sigma a(N') \bar{H}(N')\right)c^{\star}_2\left(\sigma a(N') \bar{H}(N')\right)\\  c^{\star}_1\left(\sigma a(N') \bar{H}(N')\right)c_2\left(\sigma a(N') \bar{H}(N')\right) & \left|c_2\left(\sigma a(N') \bar{H}(N')\right)\right|^2 \end{bmatrix}\,.	    	
	\end{equation} 
\end{small}

Eq. \eqref{stochastic_solution_simplified} is nothing more than the k-integral (or anti-Fourier transform) of the power spectrum coming from linear perturbation theory \eqref{power_spectrum_def}.

In fact, by using that 
\begin{small}
	\begin{align} \nonumber
		&\int_{k=\sigma a(0) \bar{H}(0)}^{k=\sigma a(N) \bar{H}(N)}\frac{k^3}{2\pi^2} \mathbf{Q}_k(N)\cdot \mathbf{Q}_k^{T}(N) \frac{dk}{k} =  \\
		& \int_{k=\sigma a(0) \bar{H}(0)}^{k=\sigma a(N) \bar{H}(N)}\frac{k^3}{2\pi^2}\exp \left[-\int_{0}^{N}\mathbf{A}(s)ds\right]\cdot\begin{bmatrix} \left|c_1\left(k\right)\right|^2 & \left(c_1\left(k\right)c^{\star}_2\left(k\right)\right)\\  \left(c^{\star}_1\left(k\right)c_2\left(k\right)\right) & \left|c_2\left(k\right)\right|^2 \end{bmatrix} \cdot\exp \left[-\int_{0}^{N}\mathbf{A}^{T}(s)ds\right]\frac{dk}{k}\,
		\label{solution_integrated_Q}\ ,
	\end{align}
\end{small}
and changing variable from $N$ to $k=\sigma a(N) \bar{H}(N)$ in \eqref{stochastic_solution_simplified}, we finally get 
\begin{equation}
	\langle \mathbf{Q}^{IR}(N) \left(\mathbf{Q}^{IR}\right)^T(N) \rangle = \int_{k=\sigma a(0) \bar{H}(0)}^{k=\sigma a(N) \bar{H}(N)}\frac{k^3}{2\pi^2} \mathbf{Q}_k(N)\cdot \mathbf{Q}_k^{T}(N) \frac{dk}{k}\,,
	\label{final_equality}
\end{equation}
which confirms our expectation. In other words, the stochastic 2-point correlator at leading order in small noise expansion is the same as the k-integral of the linear quantum field theory power spectrum, at any order in slow-roll parameters, which is also what was found numerically in \cite{Cruces:2021iwq}. 

It is important to remark here that, although the stochastic 2-point correlator from the power spectrum of linear perturbation theory is easy to find by using \eqref{final_equality}, it is not generically possible to do the other way around: because the variable of integration in \eqref{final_equality} is $\frac{dk}{k}=d\log k$, one might be tempted to try to recover the power spectrum $\frac{k^3}{2\pi^2} \mathbf{Q}_k(N)\cdot \mathbf{Q}_k^{T}(N)$ by performing the derivative of the stochastic two-point operator with respect to $\log k$ when evaluated at $k=\sigma a(N) \bar{H}(N)$, i.e.

\begin{align} \nonumber
	&\frac{d\langle\mathbf{Q}^{IR}(N) \left(\mathbf{Q}^{IR}\right)^T(N) \rangle}{d \log (\sigma a(N) \bar{H}(N))}=  \frac{1}{1-\epsilon_1}\frac{d}{dN} \langle\mathbf{Q}^{IR}(N) \left(\mathbf{Q}^{IR}\right)^T(N) \rangle \\
	&= \frac{k^3}{2\pi^2} \mathbf{Q}_k(N)\cdot \mathbf{Q}_k^{T}(N)\Bigg|_{k=\sigma a(N)\bar{H}(N)}  + \frac{1}{1-\epsilon_1}\int_{k=\sigma a(0) \bar{H}(0)}^{k=\sigma a(N) \bar{H}(N)}\frac{d}{dN}\left(\frac{k^3}{2\pi^2} \mathbf{Q}_k(N)\cdot \mathbf{Q}_k^{T}(N)\right) \frac{dk}{k}\,.
	\label{naive_derivative}	
\end{align}

Thus generically, whenever the power spectrum of the variable Q is time dependent\footnote{The power spectrum of the Mukhanov-Sasaki variable $Q_k^{UV}$ is time dependent only at $\mathcal{O}(\epsilon)$ in Slow-Roll and Ultra-Slow-Roll, during Constant-Roll the power spectrum is not constant even at zeroth order in SR parameters.}
\begin{equation} 
	\frac{1}{1-\epsilon_1}\frac{d}{dN} \langle\mathbf{Q}^{IR}(N) \left(\mathbf{Q}^{IR}\right)^T(N) \rangle \neq  \frac{k^3}{2\pi^2} \mathbf{Q}_k(N)\cdot \mathbf{Q}_k^{T}(N)\Bigg|_{k=\sigma a(N)\bar{H}(N)} \,.
\end{equation}

In other words, in stochastic inflation, we loose the scale dependence of the power spectrum. Even worse, the 2-point correlator of \eqref{final_equality} is rather arbitrary since the integration limits depends on our choice of the coarse grained scale and initial time.

It should be noted here the importance of the momentum constraint, as earlier noticed in \cite{Prokopec:2019srf}. While at zeroth order in $\epsilon_1$ the momentum constraints is trivially satisfied, at next to leading order it is not, as proved in \cite{Cruces:2021iwq}. Thus, by neglecting the momentum constraint, the equality \eqref{final_equality} would not be satisfied at all order in slow-roll parameters.

We have analytically shown the (expected) relation between the stochastic 2-point correlator and the linear power spectrum of the MS variable $Q$ at all orders in SR parameters. Previous works in this aspect usually aim to compare the stochastic correlators of the curvature perturbations (rather than of the field fluctuations) ($\mathcal{R}_{lin}$) with those computed using perturbation theory \cite{Vennin:2015hra,Firouzjahi:2018vet, Pattison:2021oen}. This can be directly done
in the linear regime where $\left(\mathcal{R}_{lin}\equiv \frac{Q_1}{\bar{\pi}}\right)$.
In this case, our results matches exactly with the literature. Nonetheless, at leading order in small noise expansion, we can go further and relate Q with $\delta N$ via gauge transformation \cite{Maldacena:2002vr}.

\section{Non-Gaussianities} \label{sec:NG}

Because the noise $\xi(N)$ is Gaussian and the stochastic system of \eqref{sto_any_epsilon0_noise2} is linear, the probability distribution related to the stochastic inflationary system, at leading order in the small noise expansion, is a bivariate Gaussian (see discussion after \eqref{small_noise}) whose covariance matrix is given by  \eqref{final_equality}. From eq. \eqref{final_equality}, we can also check that $Q^{IR}$ and $\pi_Q^{IR}$ are completely correlated, which is a direct consequence of the decoherence of quantum fluctuations at superhorizon scales, i.e. that $Q^{\star}_k\left(\pi_Q\right)_k=Q_k\left(\pi_Q\right)^{\star}_k=\pm \sqrt{\left|Q_k\right|^2}\sqrt{\left|\left(\pi_Q\right)_k\right|^2}$ \cite{Vennin:2024yzl}\footnote{Note that we have already implicitly used this condition when assuming $\xi$ in \eqref{full_stochastic_system2} to take, at a given time $N$, the same random value for all the three equations.}. This implies that, at leading order in the small noise expansion, any non-Gaussianity in the curvature perturbation might only be related to a field redefinition between $Q$ and the curvature perturbation. This is precisely what was also found via the classical $\delta N$ formalism \cite{Pi:2022ysn, Ballesteros:2024pwn},  where a Gaussian PDF for the field is generically assumed.
	
	In order to compute the intrinsic non-Gaussianity of the field itself, we must go beyond leading order in small noise expansion \cite{Ballesteros:2020sre}. As an example, we will study the SR case, where the result is known analytically \cite{Maldacena:2002vr, Allen:2005ye}. 
	
	At leading order in SR parameters (solutions at higher orders in SR parameters are non-analytical), one can show that the noises $\bar{b}_{\phi}$, $\bar{b}_{\pi}$ and $\bar{b}_{C}$ take the following form: 
	
	\begin{align} \nonumber
		\bar{b}_{\phi} = & \sqrt{1-\epsilon_1}\frac{\lambda_{\phi}}{\lambda^i_\phi}=\frac{\bar{H}(N)}{\bar{H}_0}\left[1-\frac{1}{2}\epsilon_1+\mathcal{O}\left(\epsilon^2\right)\right]\,, \\ \nonumber
		\bar{b}_{\pi} = & \sqrt{1-\epsilon_1}\frac{\lambda_{\pi}}{\lambda^i_\pi}=1+\mathcal{O}\left(\epsilon^2\right)\,, \\
		\bar{b}_{C} = & \frac{1}{\bar{\pi}}\mathcal{O}\left(\epsilon^2\right)\,,
		\label{b_SR}
	\end{align}
	where $\bar{H}_0=\bar{H}(N=0)$ and we have used that the square root of the power spectrum of the field and its velocity in SR is (see appendix \ref{app:SR} for details) 
	\begin{align} \nonumber
		\lambda_{\phi} = & \frac{\bar{H}(N)}{2\pi}\left[1+\left(\alpha-\log (\sigma)\right)\left(\epsilon_1+\frac{\epsilon_2}{2}\right)-\epsilon_1 + \mathcal{O}\left(\epsilon^2\right)\right]\,, \\ \nonumber
		\lambda_{\pi} = & \frac{\bar{H}_0}{2\pi}\left[\frac{\epsilon_2}{2}+\mathcal{O}\left(\epsilon^2\right)\right]\,, \\
		\label{lambda_SR}
	\end{align}
	where we have defined $\alpha \equiv 2 - \log(2) - \gamma$, and $\gamma$ is the Euler-Mascheroni constant. With $\mathcal{O}(\epsilon)$ we mean $\mathcal{O}(\epsilon_i) \quad \forall i$. The equations at leading order in the small noise expansion (also at leading order in SR parameters) are
	\begin{align} \nonumber
		\frac{\partial \phi_1}{\partial N} = & \pi_1 + \frac{\bar{H}(N)}{\bar{H}_0}\left[1-\frac{1}{2}\epsilon_1+\mathcal{O}\left(\epsilon^2\right)\right]\xi\,, \\ \nonumber
		\frac{\partial \pi_1}{\partial N}  + & \left(3 - \epsilon_1 + \mathcal{O}\left(\epsilon^2\right)\right)  \pi_1 + \left(-\frac{3}{2}\epsilon_2 + \mathcal{O}\left(\epsilon^2\right)\right) \phi_1 = - \xi\,, \\ 
		\bar{\pi}\frac{\partial}{\partial N}& \left(\frac{1}{3}\nabla^2 C_1\right) = \mathcal{O}\left(\epsilon^2\right) + \mathcal{O}\left(\epsilon^2\right)\xi \,.
		\label{stochastic_1st_small_1st_SR}
	\end{align}
	where we have multiplied by $\bar{\pi}$ the equation for the momentum constraint because  $\bar{\pi}\frac{\partial}{\partial N} \left(\frac{1}{3}\nabla^2 C_1\right)$ will always appear together as we can see in \eqref{def_QIR}.
	
	Note that in the last line of \eqref{stochastic_1st_small_1st_SR}, we have used the solution of $\pi_1$ and $\phi_1$ obtained by solving the closed system formed by the two first equations. During SR it is easy to show that the stochastic variables $\pi_1$ and $\phi_1$ are completely correlated and satisfy the following relation:
	\begin{equation}
		\pi_1 = \left(\frac{\epsilon_2}{2}+\mathcal{O}\left(\epsilon^2\right)\right)\phi_1\,.
		\label{phi1_pi1_SR}
	\end{equation}	
	
	For the second order in the small noise expansion we have to compute the derivatives of $\bar{b}_{\phi}$, $\bar{b}_{\pi}$ and $\bar{b}_{C}$. Using  \eqref{phi1_pi1_SR} and the background equations of motion we have
	
	\begin{equation}
		\frac{d}{d \bar{\phi}}\left(\epsilon^m\right) \phi_1 \sim \frac{d}{d \bar{\pi}}\left(\epsilon^m\right) \pi_1 \sim \frac{1}{\bar{\pi}}\epsilon^{m+1} \phi_1\,,  
	\end{equation}
	where $m$ is any integer or half-integer value larger or equal than $1$. With this in mind, the only contribution to the noises that do not vanish at leading order in SR parameters is 
	
	\begin{equation}
		\frac{d}{d \bar{\phi}}\left(\bar{b}_{\phi}\right) \phi_1  = \frac{1}{\bar{H}_0}\frac{d \bar{H}(N)}{d\bar{\phi}} +\frac{1}{\bar{\pi}}\mathcal{O}(\epsilon^2) = -\frac{\epsilon_1}{\bar{\pi}} +\frac{1}{\bar{\pi}}\mathcal{O}(\epsilon^2)\,,
		\label{b_derivative_phi}
	\end{equation}
	where we have again used \eqref{phi1_pi1_SR} to neglect $\frac{d}{d \bar{\pi}}\left(\bar{b}_{\phi}\right) \pi_1$. One can then show that, not only at second order in small noise expansion but at any order $n$ with $n\geq 2$, the small noise expansion equations are 
	
	\begin{align} \nonumber
		\frac{\partial \phi_n}{\partial N} = & \pi_n + \left(-\frac{\epsilon_1}{\bar{\pi}}+\frac{1}{\bar{\pi}}\mathcal{O}(\epsilon^2)\right)\phi_{n-1}\xi\,, \\ \nonumber
		\frac{\partial \pi_n}{\partial N}  + & \left(3 - \epsilon_1  +\frac{1}{\bar{\pi}}\mathcal{O}(\epsilon^2)\right)  \pi_n + \left(-\frac{3}{2}\epsilon_2 +\frac{1}{\bar{\pi}}\mathcal{O}(\epsilon^2)\right) \phi_n + \left(\frac{1}{\bar{\pi}}\mathcal{O}(\epsilon^2)\right)\phi_{n-1}^2= - \left(\frac{1}{\bar{\pi}}\mathcal{O}(\epsilon^3)\right)\phi_{n-1}\xi\,, \\ 
		\bar{\pi}\frac{\partial}{\partial N}& \left(\frac{1}{3}\nabla^2 C_2\right)   + \left(\frac{1}{\bar{\pi}}\mathcal{O}(\epsilon^3)\right)\phi_n +\left(\frac{1}{\bar{\pi}}\mathcal{O}(\epsilon^2)\right)\phi_{n-1}^2= \left(\frac{1}{\bar{\pi}}\mathcal{O}(\epsilon^3)\right)\phi_{n-1}\xi\,,
		\label{stochastic_2st_small_1st_SR}
	\end{align}
	
	Defining $\delta\phi_{IR}\equiv \sum_{n=1}^{\infty}\left(\lambda_{\phi}^i\right)^{n}\phi_n$ and $\delta\pi_{IR}\equiv \sum_{i=1}^{\infty}\left(\lambda_{\pi}^i\right)^{n}\pi_n$ we can re-sum all order in small noise expansion keeping only leading order in SR parameters to get the following stochastic equation
	
	\begin{align} \nonumber
		\frac{\partial \delta\phi^{IR}}{\partial N} = & \delta\pi^{IR} +\frac{\bar{H}_0}{2\pi}\left[1+\left(\alpha-\log (\sigma)\right)\left(\epsilon_1+\frac{\epsilon_2}{2}\right)-\frac{3}{2}\epsilon_1-\epsilon_1 N -\frac{\epsilon_1}{\bar{\pi}}\delta\phi^{IR} \right]\xi(N)\,, \\ 
		\frac{\partial \delta\pi^{IR}}{\partial N}  + & \left(3 - \epsilon_1\right)  \delta\pi^{IR} -\frac{3}{2}\epsilon_2  \delta\phi^{IR} = -\frac{\epsilon_2}{2}\frac{\bar{H}_0}{2\pi}\xi(N) \,, \label{stochastic_delta_1st_SR}
	\end{align}
	where we have used $\bar{H}\simeq \bar{H}_0(1-\epsilon_1 N)$. The equation for the moments of the PDF of $\delta\phi^{IR}$ described by \eqref{stochastic_delta_1st_SR} can be easily obtained by integrating by parts its corresponding Fokker-Planck equation. We leave the details of this computation in appendix \ref{app:SR} for the interested reader. 
	
	If we are now interested in typical trajectories (i.e. trajectories following the SR conditions), at zeroth order in SR parameters, the mean of $\delta\phi^{IR}$ and its variance coincide with those obtained at leading order in small noise expansion (see section \ref{sec:tree_level} and appendix \ref{app:SR}) and hence $\delta\phi^{IR}$ is Gaussian at this order. However, at next to leading order one has a SR-suppressed non-Gaussian contribution coming from the non-linear term $\delta\phi^{IR}\xi$ in \eqref{stochastic_delta_1st_SR}. This implies that the third moment of $\delta\phi^{IR}$ :	
	\begin{equation}
		\left\langle \left(\delta\phi^{IR}(N)\right)^3\right\rangle \simeq -\frac{3\epsilon_1}{\bar{\pi}}\left(\frac{\bar{H_0}}{2\pi}\right)^4N^2\,.
		\label{third_field_sol}
	\end{equation}
	
	As a check, we will show that the intrinsic non-Gaussianity of the field in \eqref{third_field_sol} is necessary to recover Maldacena's consistency condition \cite{Maldacena:2002vr} in the case in which 
	\begin{equation}
		\delta\phi^{IR} \simeq \delta\phi_g^{IR} + \frac{3}{5}f^{\phi}_{NL} \left[\left(\delta\phi^{IR}_g\right)^2-\langle\left(\delta\phi^{IR}_g\right)^2\rangle\right]\,,
		\label{ansatz_field}
	\end{equation}
	where $\delta\phi_g^{IR}$ is the first order in small noise expansion Gaussian field with zero mean and variance given by \eqref{final_equality}. Then, the 
	parameter $f_{NL}^{\phi}$ can be easily computed to be
	\begin{equation}
		f_{NL}^{\phi}\simeq \frac{5}{18}\frac{\left\langle \left(\delta\phi^{IR}(N)\right)^3\right\rangle}{\left\langle \left(\delta\phi^{IR}(N)\right)^2\right\rangle^2} = -\frac{5}{6}\frac{\epsilon_1}{\bar{\pi}}\,.
		\label{f_NL_field}
	\end{equation}
	
	In order to make connection with Maldacena's consistency condition we need to use the non-linear relation between $\mathcal{R}$ and 
	$\delta\phi$ given by the field redefinition (3.10) in \cite{Maldacena:2002vr} or by the classical $\delta N$ formalism \cite{Allen:2005ye}

	\begin{equation}
		\mathcal{R} = -\frac{\delta\phi^{IR}}{\bar{\pi}} + \frac{\epsilon_2}{4}\left[\left(\frac{\delta\phi^{IR}}{\bar{\pi}}\right)^2-\left\langle\left(\frac{\delta\phi^{IR}}{\bar{\pi}}\right)^2\right\rangle\right]
		\label{curvature_1}
	\end{equation}
	
	Inserting \eqref{ansatz_field} into \eqref{curvature_1} we have, at leading order in SR parameters,
	
	\begin{equation}
		\mathcal{R} = -\frac{\delta\phi_g^{IR}}{\bar{\pi}} + \frac{2\epsilon_1+\epsilon_2}{4}\left[\left(\frac{\delta\phi^{IR}_g}{\bar{\pi}}\right)^2-\left\langle\left(\frac{\delta\phi^{IR}_g}{\bar{\pi}}\right)^2\right\rangle\right] \equiv \mathcal{R}_g+ \frac{3}{5}f_{NL}\left[\mathcal{R}_g^2-\left\langle\mathcal{R}_g^2\right\rangle\right]\,,
		\label{curvature_2}
	\end{equation}
	where the linear comoving curvature perturbation is $\mathcal{R}_g \equiv \frac{\delta\phi^{IR}_g}{\bar{\pi}}$. 
	
	From \eqref{curvature_2} we clearly see that the parameter $f_{NL}$ satisfies Maldacena's consistency condition 
	
	\begin{equation}
		f_{NL}=\frac{5}{12}(2\epsilon_1+\epsilon_2) = -\frac{5}{12}(n_s-1)\,, 
	\end{equation}
	which confirms the validity of our small noise expansion beyond leading order.

\subsection{The inconsistency of the test field approximation and other approaches}
\label{sec:inconsistency}
In the literature, it is often stated that the test field approximation, where the IR potential of the scalar is kept fully non-linear while the noises only calculated (in our language) at leading order in $\lambda$ \cite{Starobinsky:1986fx, Nambu:1987ef, Nambu:1988je, Kandrup:1988sc, Nambu:1989uf, Mollerach:1990zf, Linde:1993xx, Casini:1998wr, Finelli:2008zg, Finelli:2010sh, Assadullahi:2016gkk, Pattison:2017mbe, Firouzjahi:2018vet, Prokopec:2019srf, Ballesteros:2020sre, Pattison:2021oen,Clesse:2015wea, Tada:2021zzj, Tomberg:2023kli, Vennin:2024yzl, Mizuguchi:2024kbl},  should capture the non-perturbative quantum behavior of the IR modes. As we have shown in this paper, this conclusion is generically incorrect. For example, intrinsic non-Gaussianities of the field only appear at next to leading order in small noise expansion in the perturbative regime.
	
	Furthermore, from \eqref{stochastic_2st_small_1st_SR} we can clearly see that the non-linearities in the IR equation (proportional to $\phi_1^2$) are even less important than the stochastic correction to the "deterministic" noises during SR\footnote{This statement is model dependent, and we have shown in appendix \eqref{app:USR_CR} that non-linearities in the IR equation are exactly of the same order in CR and USR as the stochastic correction to the "deterministic" noises}.  This proves our initial statement: it is inconsistent to consider the full IR potential while computing the noises in the deterministic fiducial background, as it is done in the test field scenario. More properly, while a stochastic system of a non-linear test field in a fixed De Sitter background makes perfect sense (see e.g. \cite{Cohen:2021fzf,Cespedes:2023aal, Palma:2023idj}), it cannot capture the non-linear behavior of the inflaton, not even at zeroth order in slow roll parameters. 

Another approach sometimes used in the literature \cite{Vennin:2015hra} is to assume that the noises computed over a stochastic background by solving \eqref{MS_equation_com_sto} have the same functional form as the noises computed over a deterministic background, but substituting the deterministic variables by stochastic ones. Specifically:	\begin{equation}
		b_{\phi} \simeq\frac{1}{\lambda_{\phi}^i} \frac{H^{IR}\left(\phi^{IR},\pi^{IR}\right)}{2\pi}\,.
		\label{Vennin_noise}
	\end{equation}
	Of course this might only be justified for typical trajectories (in the perturbative regime) but not generically (e.g. in the tail of the scalar PDF) as the \eqref{MS_equation_com_sto} does not have an analytical solution due to the stochasticity of the variables that appear therein. 
	
	Nevertheless, if we are content in recovering the intrinsic non-Gaussinity of curvature perturbation, this approach turns out to be correct \cite{Vennin:2015hra} at leading order in SR parameters: as we have shown in section \ref{sec:NG}, the actual stochastic noise in the SR case is 
	\begin{equation}
		b_{\phi} \simeq\frac{1}{\lambda_{\phi}^i} \frac{\bar{H}\left(\bar{\phi},\bar{\pi}\right)}{2\pi}\left(1+\frac{d \bar{H}}{d\bar{\phi}}\delta\phi^{IR}\right)\,,
		\label{our_noise}
	\end{equation}
	where we are neglecting the SR suppressed deterministic corrections that appear in \eqref{stochastic_delta_1st_SR} for better comparison. The noises \eqref{Vennin_noise} and \eqref{our_noise}, coincide at leading order in SR parameters. Note however that this equivalence is not guaranteed at higher orders in SR parameters or beyond SR (even for typical trajectories), where the small noise expansion should be applied consistently.

\section{Conclusions}
One of the main technical challenges that typically limits the implementation of stochastic inflation to the sole perturbative regime of cosmological perturbation, is the non-Markovian nature of the noises. This non-Markovian nature arises due to the inevitable coupling between ultra-violet and infra-red modes. What is routinely done to overcome this issue is to consider these couplings as second order in perturbation theory. However, this leads to the loss of predictive power in the non-perturbative regime of the infra-red modes. Here, we suggested a way out. While cosmological ultra-violet modes are perturbatively small under the assumption of a Bunch-Davis vacuum, infra-red modes might be non-perturbatively large due to cumulative effects from the stochastic kicks. We have implemented this idea in the realm of the small noise expansion technique. We showed that the full non-perturbative stochastic inflation can in principle be solved in terms of an infinite set of stochastic equations with Markovian noises (Wiener processes). As a bi-product we showed that, $1)$ consistently at linear order, the power spectrum always matches that of perturbation theory at all order in slow roll parameters if the momentum constraint is properly used and $2)$ the small noise approximation also recovers the known intrinsic non-Gaussianities of the field in SR, at least in the case of typical trajectories. 

Finally, we showed that the test field approximation on a De-Sitter background matches the solution of the stochastic equations at leading order in slow-roll parameters, only in the linear regime. Thus, non perturbative results of the test field approximation cannot shed any light on the non-perturbative regime of  the back-reacting inflaton field.

\acknowledgments
D.C. is supported by the National Key Research and Development Program of China Grant No.~2021YFC2203004, the National Natural Science Foundation of China (NSFC) Grant No. 12475066 and since the 01/01/2025 by the NSFC Grant No. 12447160. C.G. would like to thank Gerasimos Rigopoulos for useful discussions on the IR-UV mode separation. CG is supported by the grants number 2021-SGR-872 and PID2022-136224NB-C22. A.N. would like to express gratitude to the Institute for Research in Fundamental Sciences (IPM) and the Institute of Science Tokyo for their hospitality and support during the course of this work. Additionally, appreciation is extended to Iran's National Elites Foundation for their financial assistance.  M.Y. is supported by IBS under the project code, IBS-R018-D3, and by JSPS Grant-in-Aid for Scientific Research Number JP21H01080. We would like to thank the anonymous referee for the useful comments on our previous draft.

\appendix
\section{Stochastic equations in ultra-slow-roll and constant-roll}
\label{app:USR_CR}

In this appendix we shall write down the stochastic equations up to order $\lambda^2$ in the Ultra and Constant- roll inflationary non-attractor evolutions.
\subsection{Ultra Slow Roll Inflation (USR)}
At zeroth order in $\mathcal{O}(\epsilon_1)$ (where we can write an analytical solution), the noises $b_{\phi}$, $b_{\pi}$ and $b_{C}$ are:

\begin{align} \nonumber
	\bar{b}_{\phi} = & \frac{1}{\lambda}\frac{\bar{H}_0}{2\pi}\left[1+ \mathcal{O}\left(\frac{\epsilon_1}{\bar{\pi}}\right)\right]\,, \\ \nonumber
	\bar{b}_{\pi} = & \frac{1}{\lambda}\frac{\bar{H}_0}{2\pi}\left[\mathcal{O}\left(\frac{\epsilon_1}{\bar{\pi}}\right)\right]\,, \\
	\bar{b}_{C} = & \mathcal{O}\left(\frac{\epsilon_1}{\bar{\pi}}\right)\,.
	\label{b_USR}
\end{align}
Note that in \eqref{b_USR} we have written the corrections to be of order $\mathcal{O}\left(\frac{\epsilon_1}{\bar{\pi}}\right)$ instead of of order $\mathcal{O}\left(\epsilon_1\right)$. The reason is that, during USR, the decaying mode, which is proportional to $a^{-3}\sim \bar{\pi} \sim \frac{\epsilon_1}{\bar{\pi}}$ is actually more important than the correction at order $\mathcal{O}\left(\epsilon_1\right)$ 

By using that, for USR:
\begin{align} \nonumber
	\epsilon_i^{USR}=-6+2\epsilon_1^{USR} & \qquad \text{when} \quad i \quad \text{even}. \\
	\epsilon_i^{USR}=2\epsilon_1^{USR} & \qquad\text{when} \quad i>1 \quad \text{and odd}.
	\label{epsilon_USR}
\end{align}
we get
\begin{align} \nonumber
	\frac{\partial \phi_1}{\partial N} = & \pi_1 + \frac{1}{\lambda}\frac{\bar{H}_0}{2\pi}\left[1+ \mathcal{O}\left(\frac{\epsilon_1}{\bar{\pi}}\right)\right]\xi\,, \\ \nonumber
	\frac{\partial \pi_1}{\partial N}  + & \left(3 + \mathcal{O}\left(\epsilon_1\right)\right)  \pi_1 = - \frac{1}{\lambda}\frac{\bar{H}_0}{2\pi}\left[\mathcal{O}\left(\frac{\epsilon_1}{\bar{\pi}}\right)\right]\xi\,, \\ 
	\frac{\partial}{\partial N} & \left(\frac{1}{3}\nabla^2 C_1\right) = \mathcal{O}\left(\frac{\epsilon_1}{\bar{\pi}}\right)\,.
	\label{stochastic_1st_small_1st_USR}
\end{align}
From \eqref{stochastic_1st_small_1st_USR} we have a relation between the stochastic variables $\pi_1$ and $\phi_1$ during USR such that
\begin{equation}
	\pi_1 = \mathcal{O}\left(\frac{\epsilon_1}{\bar{\pi}}\right)\phi_1\,,
	\label{phi1_pi1_USR}
\end{equation}
Taking into account that $\frac{d \epsilon_1}{d \bar{\phi}} = 0$ and $\frac{d \epsilon_1}{d \bar{\pi}} = 2\frac{\epsilon_1}{\bar{\pi}}$ and the relation \eqref{epsilon_USR}, we can finally write
\begin{align} \nonumber
	\frac{\partial \phi_2}{\partial N} = & \pi_2 + \frac{1}{\lambda}\frac{\bar{H}_0}{2\pi}\left[\mathcal{O}\left(1\right)\right]\pi_1\xi\,, \\ \nonumber
	\frac{\partial \pi_1}{\partial N}  + & \left(3 + \mathcal{O}\left(\epsilon_1\right)\right)  \pi_1  - \frac{6\epsilon_1}{\bar{\pi}}\pi_1^2= - \frac{1}{\lambda}\frac{\bar{H}_0}{2\pi}\left[\mathcal{O}\left(1\right)\right]\xi\,, \\ 
	\frac{\partial}{\partial N} & \left(\frac{1}{3}\nabla^2 C_1\right) + \frac{\epsilon_1}{(3-\epsilon_1)\bar{\pi}}\pi_2 - \frac{\epsilon_1}{\bar{\pi}}\phi_2 + \frac{\pi_1\phi_1}{2M_{PL}^2} + \frac{3}{4(3-\epsilon_1)^2 M_{PL}^2}\pi_1^2 = \mathcal{O}\left(1\right)\pi_1 \xi\,,
	\label{stochastic_2nd_small_1st_USR_1}
\end{align}
or	
\begin{align} \nonumber
	\frac{\partial \phi_2}{\partial N} = & \pi_2 + \frac{1}{\lambda}\frac{\bar{H}_0}{2\pi}\left[\mathcal{O}\left(\frac{\epsilon_1}{\bar{\pi}}\right)\right]\phi_1\xi\,, \\ \nonumber
	\frac{\partial \pi_1}{\partial N}  + & \left(3 + \mathcal{O}\left(\epsilon_1\right)\right)  \pi_1  - \mathcal{O}\left(\frac{\epsilon_1}{\bar{\pi}}\right) \epsilon_1 \phi_1^2= - \frac{1}{\lambda}\frac{\bar{H}_0}{2\pi}\left[\mathcal{O}\left(\frac{\epsilon_1}{\bar{\pi}}\right)\right]\phi_1\xi\,, \\ 
	\frac{\partial}{\partial N} & \left(\frac{1}{3}\nabla^2 C_1\right) +  \mathcal{O}\left(\frac{\epsilon_1}{\bar{\pi}}\right)\left(1+\mathcal{O}\left(\frac{\epsilon_1}{\bar{\pi}}\right)\right)\left(\phi_2+\phi_1^2\right) = \mathcal{O}\left(\frac{\epsilon_1}{\bar{\pi}}\right)\phi_1 \xi\,.
	\label{stochastic_2nd_small_1st_USR_1}
\end{align}
 We see that the quadratic corrections in the IR equation of motion are of the same order as the stochastic correction to the deterministic noises. This means that, if we want to include decaying terms into our stochastic system \eqref{full_stochastic_system}, we must compute the noises over the stochastically corrected background. On the other hand, if we neglect decaying terms, the stochastic system at all orders in small noise expansion turns out to be:
\begin{align} \nonumber
	\frac{\partial \phi^{IR}}{\partial N} = & \pi^{IR} + \frac{\bar{H}_0}{2\pi}\xi\,, \\ \nonumber
	\frac{\partial \pi^{IR}}{\partial N}  + &  3\pi^{IR} = 0\,.
\end{align}
\subsection{Constant Roll inflation (CR)}
In this case, we define $\kappa \equiv \frac{V_{\phi}(\bar{\phi})}{\bar{H}^2\bar{\pi}}$ to be exactly a constant. We will not repeat the computation here again because it follows the same steps as in SR and USR.  The conclusions are the same as in USR i.e.,
the quadratic corrections in the IR equation of motion are of the same order as the stochastic correction to the deterministic noises (in this case of order $\mathcal{O}\left(\frac{\epsilon_1}{\bar{\pi}}\right)$). If we are willing to neglect $\mathcal{O}\left(\frac{\epsilon_1}{\bar{\pi}}\right)$ terms, then the stochastic system at all orders in the small noise approximation is
\begin{align} \nonumber
	\frac{\partial\phi^{IR}}{\partial N} & = \pi^{IR} + \frac{\bar{H}_0}{2 \pi}\frac{\Gamma\left[\nu_{CR}\right]}{\Gamma\left[\frac{3}{2}\right]}\left(\frac{\sigma}{2}\right)^{\frac{3}{2}-\nu_{CR}}\xi\,,\\
	\frac{\partial \pi_1^{IR}}{\partial N} & + 3\pi_1^{IR}-\kappa(3+\kappa)\phi_1^{IR} = -  \left(3+\kappa\right)\frac{\bar{H}_0}{2 \pi}\frac{\Gamma\left[\nu_{CR}\right]}{\Gamma\left[\frac{3}{2}\right]}\left(\frac{\sigma}{2}\right)^{\frac{3}{2}-\nu_{CR}}\xi\,,		\end{align}
 where $\nu_{CR} = \frac{3}{2}\left|1+\frac{2}{3}\kappa\right|$. Note that, for $\kappa>0$ or $\kappa<-3$, the condition for the small noise expansion eventually breaks down as the power spectrum of the scalar field exponentially grows in time  and hence the noises become large for $\sigma \ll 1$.

Again, if one wish to include $\mathcal{O}\left(\frac{\epsilon_1}{\bar{\pi}}\right)$ effects in CR, which appears as quadratic terms in the IR stochastic variables, then one must also take into account the stochastic corrections to the deterministic noises.

\section{Detailed solution in SR}
\label{app:SR}

As an example, in this appendix we will explicitly solve \eqref{stochastic_delta_1st_SR} and show that the variance of the probability distribution for the field up to leading order in SR parameters is given by \eqref{final_equality}.

In order to compute $\bar{b}_{\phi}$, $\bar{b}_{\pi}$ and $\bar{b}_{C}$ we should solve the linearized Einstein equations in uniform-N gauge. However, at leading order in SR parameters, uniform-N gauge and flat gauge are equivalent (at least at superhorizon scales) so for simplicity we will solve the linearized Einstein equation in flat gauge (see \cite{Cruces:2022imf, Pattison:2019hef, Figueroa:2021zah} for discussions about these two gauges in the context of stochastic inflation), which is equivalent to solve the MS equation \eqref{MS_equation_Q_det}. Its solution at superhorizon scales during SR, by the use of Bunch-Davies initial conditions \cite{Bunch:1978yq} is (see for example \cite{Cruces:2018cvq}):

\begin{equation}
	Q^{UV}_k \simeq \frac{(1-i)2^{-\frac{3}{2}+\nu}e^{\frac{1}{2} i \pi \nu} \sqrt{-\tau} \Gamma[\nu]}{a\sqrt{\pi}}\left(-k\tau\right)^{-\nu}
	\label{sol_MS_SR}\,,
\end{equation}
where $\tau\simeq \frac{1}{a \bar{H}}\left(1+\epsilon_1\right)$ is the conformal time and $\nu=\frac{3}{2}+\epsilon_1+\frac{\epsilon_2}{2}$ during SR.

With solution \eqref{sol_MS_SR} we can easily compute \eqref{lambda_SR} and hence, following the procedure explained in the main text, one can straightforwardly arrive at \eqref{stochastic_delta_1st_SR}, which is the equation that we want to solve.

Now, we can diagonalize the system \eqref{stochastic_delta_1st_SR} by defining the following variables:
	
	\begin{align} \nonumber
		\tilde{\delta\phi}^{IR}\equiv\delta\phi^{IR}+\frac{\epsilon_2}{2}\delta\pi^{IR}\,, & \qquad & \tilde{\delta\pi}^{IR}\equiv \delta\pi^{IR}-\frac{\epsilon_2}{2}\delta\phi^{IR}\,, \\
		\tilde{\lambda}_{\phi}\equiv \lambda_{\phi}+\frac{\epsilon_2}{2}\lambda_{\pi} \simeq \lambda_{\phi} + \mathcal{O}(\epsilon^2)\,, & \qquad & \tilde{\lambda}_{\pi}\equiv \lambda_{\pi}-\frac{\epsilon_2}{2}\lambda_{\phi} \simeq \mathcal{O}(\epsilon^2)\,.
		\label{new_variables}
	\end{align}
	
	In terms of these new variables, the system \eqref{stochastic_delta_1st_SR} becomes
	
	\begin{align} \nonumber
		\frac{\partial \tilde{\delta\phi}^{IR}}{\partial N} = & \left(1-\frac{3}{2}\epsilon_2\right)\tilde{\delta\pi}^{IR} + \frac{\epsilon_2}{2}\tilde{\delta\phi}^{IR}\\ \nonumber & +  \frac{\bar{H}_0}{2\pi}\left[1+\left(\alpha-\log (\sigma)\right)\left(\epsilon_1+\frac{\epsilon_2}{2}\right)-\frac{3}{2}\epsilon_1-\epsilon_1 N -\frac{\epsilon_1}{\bar{\pi}}\delta\phi^{IR} \right]\xi(N)\,, \\
		\frac{\partial \tilde{\delta\pi}^{IR}}{\partial N}  = & - \left(3-\epsilon_1+\frac{\epsilon_2}{2}\right)\tilde{\delta\pi}^{IR}\,.
		\label{stochastic_SR_new}
	\end{align}
	
	In order to solve \eqref{stochastic_SR_new}, we must give some initial conditions at the initial time $N=0$. Since $\delta\phi^{IR}$ and $\delta\pi^{IR}$ are fluctuations, we will give as initial conditions its value when the stochastic simulation starts and everything is deterministic i.e. $\tilde{\delta\phi}^{IR}=0$ and $\tilde{\delta\pi}^{IR}=0$. Because of the second equation in \eqref{stochastic_SR_new}, we have $\tilde{\delta\pi}^{IR}\simeq 0$ always. This further simplifies the stochastic system to a very simple stochastic equation
	
	\begin{equation}
		\frac{\partial \delta\tilde{\phi}^{IR}}{\partial N} = \frac{\epsilon_2}{2}\tilde{\delta\phi}^{IR} +\frac{\bar{H}_0}{2\pi}\left[1+\left(\alpha-\log (\sigma)\right)\left(\epsilon_1+\frac{\epsilon_2}{2}\right)-\frac{3}{2}\epsilon_1-\epsilon_1 N -\frac{\epsilon_1}{\bar{\pi}}\delta\phi^{IR} \right]\xi(N)\,.
	\end{equation}
	
	Finally, note that $\tilde{\delta\pi}^{IR}\simeq 0$ implies that $\delta\pi^{IR}\simeq \frac{\epsilon_2}{2}\delta\phi^{IR}$ and hence $\tilde{\delta\phi}^{IR}\simeq \delta\phi^{IR}$, which means that the equation above is also valid for the original variable $\delta\phi^{IR}$ at leading order in SR parameters. The final equation to solve is 
	\begin{equation}
		\frac{\partial \delta\phi^{IR}}{\partial N} = \frac{\epsilon_2}{2}\delta\phi^{IR} +\frac{\bar{H}_0}{2\pi}\left[1+\left(\alpha-\log (\sigma)\right)\left(\epsilon_1+\frac{\epsilon_2}{2}\right)-\frac{3}{2}\epsilon_1-\epsilon_1 N -\frac{\epsilon_1}{\bar{\pi}}\delta\phi^{IR} \right]\xi(N)\,.
		\label{stochastic_simple}
	\end{equation}
	
	We will start by writing the Fokker-Planck  equation for the probability distribution $P(\delta\phi^{IR}, N)$ of $\delta\phi^{IR}$ at a given time $N$ associated to the Langevin equation \eqref{stochastic_simple}. At leading order in SR parameters this is
	
	\begin{align} \nonumber
		&\frac{\partial P(\delta\phi^{IR}; N)}{\partial N}  =  -\frac{\epsilon_2}{2}\frac{\partial}{\partial \delta\phi^{IR}}\left[\delta\phi^{IR}P(\delta\phi^{IR}; N)\right]\\ & + \frac{1}{2}\frac{\partial^2 }{\partial \left(\delta\phi^{IR}\right)^2}\left[\left(\frac{\bar{H}_0}{2\pi}\right)^2\left[1+2\left(\alpha-\log (\sigma)\right)\left(\epsilon_1+\frac{\epsilon_2}{2}\right)-3\epsilon_1-2\epsilon_1 N -2\frac{\epsilon_1}{\bar{\pi}}\delta\phi^{IR} \right]P(\delta\phi^{IR};N)\right]\,.
		\label{FP_field}
	\end{align}
	
	We can now multiply both sides of the equation by $\left(\delta{\phi}^{IR}\right)^n$ and integrate by parts to obtain a equation for the moments of $P(\delta{\phi}^{IR};N)$, which at leading order in SR parameters is
	
	\begin{align} \nonumber
		\frac{\partial \left\langle \left(\delta\phi^{IR}\right)^n\right\rangle}{\partial N} & =  n\frac{\epsilon_2}{2}\left\langle \left(\delta\phi^{IR}\right)^n\right\rangle - n(n-1)\left(\frac{\bar{H}_0}{2\pi}\right)^2\left(\frac{\epsilon_1}{\bar{\pi}}\right)\left\langle \left(\delta\phi^{IR}\right)^{n-1}\right\rangle\\ 
		& +  \frac{n(n-1)}{2}\left(\frac{\bar{H}_0}{2\pi}\right)^2\left[1+2\left(\alpha-\log (\sigma)\right)\left(\epsilon_1+\frac{\epsilon_2}{2}\right)-3\epsilon_1-2\epsilon_1 N\right]\left\langle \left(\delta\phi^{IR}\right)^{n-2}\right\rangle\,.
		\label{moments_field}
	\end{align}
	
	We will now solve \eqref{moments_field} up to $n=3$ with initial conditions $\left\langle \left(\delta\phi^{IR}(N=0)\right)^n\right\rangle=0$ to compute the non-gaussianity.
	
	\begin{itemize}
		\item $n=1$
		\begin{align} 
			\frac{\partial \left\langle \delta\phi^{IR}\right\rangle}{\partial N}=& \langle \delta\phi^{IR}\rangle\frac{\epsilon_2}{2} \quad \rightarrow \quad  \left\langle \delta\phi^{IR}(N)\right\rangle=0\,.
			\label{mean_field}
		\end{align}
		
		\item $n=2$
		\begin{align} \nonumber
			\frac{\partial \left\langle \left(\delta\phi^{IR}\right)^2\right\rangle}{\partial N}=& \epsilon_2\left\langle \left(\delta\phi^{IR}\right)^2\right\rangle
			\\ + & \left(\frac{\bar{H}_0}{2\pi}\right)^2\left[1+2\left(\alpha-\log (\sigma)\right)\left(\epsilon_1+\frac{\epsilon_2}{2}\right)-3\epsilon_1-2\epsilon_1 N\right]\,,
			\label{variance_field_eq}
		\end{align}
		where we have used that $\left\langle\delta\phi^{IR}\right\rangle =0$ and $\left\langle\left(\delta\phi^{IR}\right)^0\right\rangle=1$. The solution to \eqref{variance_field_eq} at leading order in SR parameters is
		\begin{align} 
			\left\langle \left(\delta\phi^{IR}(N)\right)^2\right\rangle\simeq \frac{\bar{H}_0^2}{4\pi^2}N\left[1+2\left(\alpha-\log(\sigma)\right)\left(\epsilon_1+\frac{\epsilon_2}{2}\right)-3\epsilon_1 - N\left(\epsilon_1-\frac{\epsilon_2}{2}\right)\right]\,.
			\label{variance_field_sol}
		\end{align}
		
		\item $n=3$
		\begin{align}
			\frac{\partial \left\langle \left(\delta\phi^{IR}\right)^3\right\rangle}{\partial N}=& \frac{3}{2}\epsilon_2\left\langle \left(\delta\phi^{IR}\right)^3\right\rangle-6 \left(\frac{\bar{H}_0}{2\pi}\right)^4 N\left(\frac{\epsilon_1}{\bar{\pi}}\right) \,,
			\label{third_field_eq}
		\end{align}
		where we have again used that  $\left\langle\delta\phi^{IR}\right\rangle =0$, together with the solution for $\left\langle\left(\delta\phi^{IR}\right)^2\right\rangle$ given by \eqref{variance_field_sol}. The solution of \eqref{third_field_eq}
		at leading order in SR parameters is
		\begin{equation}
			\left\langle \left(\delta\phi^{IR}(N)\right)^3\right\rangle \simeq -\frac{3\epsilon_1}{\bar{\pi}}\left(\frac{\bar{H_0}}{2\pi}\right)^4N^2\,,
			\label{third_field_sol_app}
		\end{equation}
		which is the result written in the main text.
	\end{itemize}

\subsection{Comparison of the stochastic variance with linear perturbation theory.}
We will now compare \eqref{variance_field_sol} with what we would expect from linear perturbation theory. In order to make a meaningful comparison, we must compute the two point correlator of the field in real space, which is defined as

\begin{equation}
	\langle 0|\hat{\delta \phi}\left(\textbf{x}_1, N_1\right)\hat{\delta \phi}\left(\textbf{x}_2, N_2\right)|0\rangle = \int \frac{d\textbf{k}_1 d\textbf{k}_2}{\left(2\pi\right)^3}\langle 0|\hat{\delta \phi}_{\textbf{k}_1}\left( N_1\right)\hat{\delta \phi}_{\textbf{k}_2}\left( N_2\right)|0\rangle\\,,
	\label{quantum_correlator}
\end{equation}
where $\hat{\delta \phi}_{\textbf{k}}\left( N\right)=\phi_k(N)\hat{a}_{\textbf{k}}e^{i \textbf{k}\cdot \textbf{x}} + \phi^{*}_k(N)\hat{a}^{\dagger}_{\textbf{k}}e^{-i \textbf{k}\cdot \textbf{x}}$ is a quantum operator ($\hat{a}_{\textbf{k}}$ and $\hat{a}^{\dagger}_{\textbf{k}}$ are the typical creation and annihilation operators) and $\phi_k$ is the solution of the MS equation in Fourier space \eqref{MS_equation_Q_det}.

By inserting the definition of $\hat{\delta \phi}_{\textbf{k}}\left( N\right)$ into \eqref{quantum_correlator} we get 

\begin{equation}
	\langle 0|\hat{\delta \phi}\left(\textbf{x}_1, N_1\right)\hat{\delta \phi}\left(\textbf{x}_2, N_2\right)|0\rangle = \int_0^{\infty} \frac{d k}{2\pi^2}k^2 \phi_k(N_1)\phi_k^{*}(N_2)\frac{\sin k r}{k r}\,,
\end{equation}
where we have used spherical coordinates.

The stochastic formalism only computes the probability distribution of $\phi$ at a given time $N_1=N_2=N$ and at a given space-point $\textbf{x}_1=\textbf{x}_2 \rightarrow r=0$. Furthermore, it only takes into account the modes that enter the coarse-grained scale from $N=0$ onwards. With this in mind, the real space 2-point correlator that must be compared with the stochastic variance \eqref{variance_field_sol} is the following 

\begin{equation}
	\langle \delta \phi_{lin}^2\rangle=\int_{k=\sigma a(0)H(0)}^{k=\sigma a(N) H(N)}\frac{dk}{2\pi^2}k^2\left|\phi_k(N)\right|^2=\int_{k=\sigma H_0}^{k=\sigma e^N H(N)}\frac{dk}{k}\mathcal{P}_{\delta\phi}(k,N)\,,
	\label{variance_lin_field_def}
\end{equation} 
where we have used that $a(N)=e^N$ and we have defined the dimensionless power spectrum of $\delta\phi$ as

\begin{equation}
	\mathcal{P}_{\delta\phi}(k,N)\equiv \frac{k^3}{2\pi^2}\left|\phi_k(N)\right|^2\,.
\end{equation}

$\mathcal{P}_{\delta\phi}(k,N)$ can be easily computed with \eqref{sol_MS_SR} in SR, reading 

\begin{equation}
	\mathcal{P}_{\delta\phi}(k,N)=\frac{\bar{H}(N)^2}{4\pi^2}\left[1+2\left(\alpha-\log\left(\frac{k}{e^N \bar{H}(N)}\right)\right)\left(\epsilon_1+\frac{\epsilon_2}{2}\right)-2\epsilon_1\right] + \mathcal{O}(\epsilon^2)\,.
	\label{PS_field}
\end{equation}

Inserting \eqref{PS_field} into \eqref{variance_lin_field_def} we get 

\begin{equation}
	\langle \delta \phi_{lin}^2\rangle= \frac{\bar{H}_0^2}{4\pi^2}N\left[1+2\left(\alpha-\log(\sigma)\right)\left(\epsilon_1+\frac{\epsilon_2}{2}\right)-3\epsilon_1 - N\left(\epsilon_1-\frac{\epsilon_2}{2}\right)\right] + \mathcal{O}(\epsilon^2)\,,
	\label{variance_lin_field}
\end{equation}
where we have again expanded $H(N)\simeq H_0(1-\epsilon_1 N)$. We can clearly see that \eqref{variance_lin_field} is exactly the same expression as \eqref{variance_field_sol}. As we have discussed in the main text, this equality goes beyond the $\mathcal{O}(\epsilon^2)$ approximation  at leading order in small noise expansion and indeed it is valid at all orders in SR parameters.

\bibliography{references-small-noise}
\bibliographystyle{ieeetr}



\end{document}